\documentclass[aps,prb,amsmath,amssymb,reprint,superscriptaddress]{revtex4-1}
\usepackage{graphicx}
\usepackage{dcolumn}
\usepackage{bm}
\usepackage{multirow}
\usepackage{color}
\usepackage{cases}
\usepackage{ulem}

\begin{document}

\title{Hybrid Electron Spin Resonance and Whispering Gallery Mode Resonance Spectroscopy of Fe$^{3+}$ in Sapphire}

\author{Karim Benmessai}
\email{karim.benmessai@uwa.edu.au}
\affiliation{ARC Centre of Excellence for Engineered Quantum Systems, University of Western Australia, 35 Stirling Highway, Crawley WA 6009, Australia} 

\author{Warrick G. Farr}
\affiliation{ARC Centre of Excellence for Engineered Quantum Systems, University of Western Australia, 35 Stirling Highway, Crawley WA 6009, Australia} 

\author{Daniel L. Creedon}
\affiliation{ARC Centre of Excellence for Engineered Quantum Systems, University of Western Australia, 35 Stirling Highway, Crawley WA 6009, Australia} 

\author{Yarema Reshitnyk}
\affiliation{ARC Centre of Excellence for Engineered Quantum Systems, University of Queensland, St Lucia QLD 4067, Australia} 

\author{Jean-Michel Le Floch}
\affiliation{ARC Centre of Excellence for Engineered Quantum Systems, University of Western Australia, 35 Stirling Highway, Crawley WA 6009, Australia} 

\author{Timothy Duty}
\affiliation{ARC Centre of Excellence for Engineered Quantum Systems, University of New South Wales, Sydney NSW 2052, Australia} 

\author{Michael E. Tobar}
\affiliation{ARC Centre of Excellence for Engineered Quantum Systems, University of Western Australia, 35 Stirling Highway, Crawley WA 6009, Australia} 
\date{\today}
\begin{abstract}
The development of a new era of quantum devices requires an understanding of how paramagnetic dopants or impurity spins behave in crystal hosts. Here, we describe a new spectroscopic technique which uses traditional Electron Spin Resonance (ESR) combined with the measurement of a large population of electromagnetic Whispering Gallery (WG) modes. This allows the characterization of the physical parameters of paramagnetic impurity ions in the crystal at low temperatures. We present measurements of two ultra-high purity sapphires cooled to 20 mK in temperature, and determine the concentration of Fe$^{3+}$ ions and their frequency sensitivity to a DC magnetic field. Our method is different to ESR in that it is possible to track the resonant frequency of the ion from zero applied magnetic field to any arbitrary value, allowing excellent measurement precision. This high precision reveals anisotropic behaviour of the Zeeman splitting, which has not been previously reported. In both crystals, each Zeeman component demonstrates a different $g$-factor.
\end{abstract}

\pacs{Valid PACS appear here}

\maketitle

Since the development of the first Whispering Gallery maser oscillator based on Fe$^{3+}$ ions in sapphire, the system has been the focus of research at several institutions.\cite{Pyb2005APL, Benmessai2005EL, Pyb2006IJMP,  Benmessai2008PRL, Creedon2009IEEE, Benmessai2009, Karim2010, Creedon2012, Creedon2012comb} However, the physical parameters of these paramagnetic ions in sapphire have never been characterised with sufficient accuracy. Several experiments have attempted to measure the spin-spin relaxation time, or the number of ions involved in the maser process, but have given results which contradict the values given by the manufacturer of the crystals, or those found in the literature.\cite{Bogle1958, Bogle1959, Symmons1962, Kornienko1961} This difference is essentially due to the method used to interact with the ions in sapphire, namely the excitation of high-$Q$ Whispering Gallery modes at frequencies coincident with the ESR frequency of the ion. The spatial distribution of these modes acts as filter, selecting only the population of ions which fall inside its field pattern. Furthermore, the inhomogeneous broadening of the ion's ESR bandwidth makes the characterization of their physical parameters difficult. Until now, values in the literature have not been confirmed by another method.\cite{mathese, Creedon2009IEEE} The present work describes a new technique for measuring these parameters, which in principle is valid at any temperature. In our work, crystals are cooled to near 20 mK in a dilution refrigerator, and all WG modes within a few GHz of the ESR frequency of Fe$^{3+}$ in sapphire (12.04 GHz) are identified. Then, a magnetic field is applied parallel to the cylindrical $z$-axis of the crystal, and swept in value. At each value of magnetic field, the reflection and transmission coefficients of the WG mode population are measured. Characterisation of the frequency and quality factor of the modes allows the ESR bandwidth and the concentration of the ions to be determined accurately enough to confirm the work published previously. This step is very important for such a system, because it is an excellent candidate for use in a host of quantum devices, with sapphire typically containing many species of residual paramagnetic spins compared to the current engineered quantum systems. Due to the inhomogeneous broadening effect, different spin packets are dispersed to different frequencies. It has been shown that these may act independently, with a coupling between the spin packets governed by the hyperfine interaction with the aluminium in sapphire lattice\cite{Creedon2012}. These `self' interactions have also been observed exhibit non-linear behaviour. Combining this with the high $Q$-factors of the WG modes, sapphire is an attractive material for future quantum devices.\cite{Amsuss2011, Wirth2010, Bushev2011} It is therefore important to fully characterize the physical parameters of any ion defect centers in the crystals.


\section{Ion description}

The fundamental energy levels of the Fe$^{3+}$ ion in sapphire are defined by the zero-field splitting parameter $^6S$ in the spin Hamiltonian.\cite{HamiltonianBleaney, Symmons1962, Kornienko1961, Buzare2002} Three levels exist at zero applied magnetic field, with the dependence of these levels on a DC magnetic field described by the following Hamiltonian:\nopagebreak
\begin{multline}
\textrm{\bf{H}}=g\mu_B B \textrm{\bf{S}}+D\left[ \textrm{\bf{S}}_z^2-\dfrac{1}{3} S(S+1)\right] \\
+\dfrac{1}{6}a\left[ \textrm{\bf{S}}_\xi^4+\textrm{\bf{S}}_\eta^4+\textrm{\bf{S}}_\zeta^4 \right] -\dfrac{1}{5}\left[S(S+1)(3S^2+3S-1)\right]\\
+ \dfrac{1}{180}F\left[ 35\textrm{\bf{S}}_z^4 -30S(S+1)\textrm{\bf{S}}_z^2+25\textrm{\bf{S}}_z^2-6S(S+1)\right.\\
+\left.3S^2(S+1)^2\right]
\end{multline}

where $g\approx2$ is the Land\'e $g$-factor, $\mu_B=9.274\times10^{-24}$ J T$^{-1}$ is the Bohr magneton, $B$ is the DC magnetic field strength and $S$ is the electron spin angular momentum of the ion. The Hamiltonian parameters are summarized in Table \ref{tab:Kornienko}. These parameters were measured at 4.2K during the 1960s and vary from crystal to crystal due to different spatial distribution and concentration of ions in the lattice, and due to crystal dislocations and variations in the symmetry of the crystalline field. Here, our calculations use the parameters given by Kornienko and Prokhorov as they give results close to those measured in our crystals. The Hamiltonian parameters have never been calculated for extremely low temperatures, near 20 mK. Nevertheless, we assume in our calculations that they do not change significantly from their values near 4 K, especially the $g$-factor,  transition frequencies, and sensitivity to DC magnetic field.
\begin{table}[h!]
\begin{tabular}{lcccr} 
 & & Symmons \hspace{20pt} & & Kornienko\\ \hline \hline
$D$ \hspace{40pt} & & 1719.2 $\pm$ 1 \hspace{20pt} & & 1838.5 $\pm$ 0.6 \\
$|a|$ & & 229.4 $ \pm$ 1 \hspace{20pt} & & 253.5 $\pm$ 1.3 \\
$a-F$ & & 351.5 $\pm$ 1 \hspace{20pt} & & 362.7 $\pm$ 2 \\ \hline
\end{tabular}
\caption{\label{tab:Kornienko}Hamiltonian parameters for Fe$^{3+}$ in sapphire, given by Kornienko \& Prokhorov and Symmons \& Bogle at 4.2 K in units of 10$^{-4}$ cm$^{-1}$.\\}
\end{table}

\begin{figure}[h!!!!!!]
\includegraphics[width=8cm]{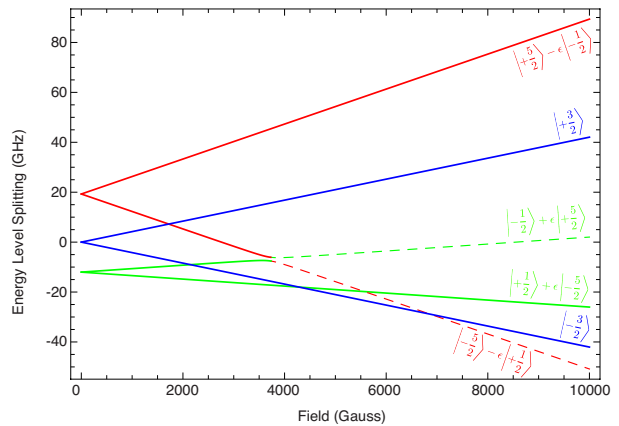}
\caption{\label{fig:energylevels}Zeeman splitting of the ESR energy levels of Fe$^{3+}$ in sapphire in the presence of a DC magnetic field. The presence of the sapphire cubic field, represented by the $a$ term in the Hamiltonian, creates a small admixing between the $\left|\pm5/2\rangle\right.$ and $\left|\mp 1/2\rangle\right.$ states ($\epsilon \simeq 0.03$)\cite{Symmons1962}.}
\end{figure}

Zeeman splitting occurs in the presence of a DC magnetic field, with each energy level split into two sub-levels as shown in Fig. \ref{fig:energylevels}. Each sub-level has a sensitivity different to the others, and crossing of sub-levels occurs at large values of magnetic field. Allowed transitions between sub-levels follow the selection rule $\Delta m=\pm1$. At zero magnetic field the ions occupying the first energy level will split in to two distinct components; one which we denote ESR$_{\text{low}}$ corresponding to the transition $\left|+1/2\rangle\right. \rightarrow \left|+3/2\rangle\right.$, where the frequency or energy difference decreases when $B$ increases, and the other which we denote ESR$_{\text{high}}$. This corresponds to the transition $\left|-1/2\rangle\right. \rightarrow \left|-3/2\rangle\right.$ where the frequency or energy difference increases with $B$.

In this work, we examine the transition between the lowest energy levels. The choice of these transitions i justified by the setup condition (Section \ref{experiment}), and because nonlinear behaviour has previously been observed at their frequencies. The energy difference is expressed for the two transitions as a function of the DC magnetic field strength $B$ (expressed in Tesla) as follows:

\begin{multline}\label{eqn:energy}
\Delta E_{\left|\pm1/2\rangle\right. \rightarrow \left|\pm3/2\rangle\right.} (B) =\pm B-D + \dfrac{3}{2}(a-F)\\
+\dfrac{1}{6}\sqrt{(9B+18D+(a-F))^2+80a^2}
\end{multline}
and the corresponding transition frequencies are expressed as
\begin{equation}\label{eqn:fre}
\nu_{\text{high/low}}=\dfrac{g \mu_B}{h}\Delta E_{\left|\pm1/2\rangle\right. \rightarrow \left|\pm3/2\rangle\right.}.
\end{equation}

\begin{figure}[h!!!!!!]
\centering
\includegraphics[width=9cm]{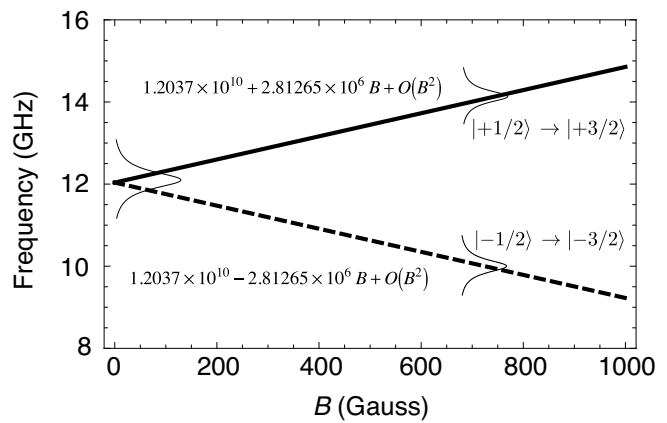}
\caption{\label{fig:deltanu}Evolution of the frequency of the transition between Zeeman levels with $B$.}
\end{figure}

The predicted evolution of the transition frequency between the states is shown in Fig. \ref{fig:deltanu}.


\section{Experimental setup}
\label{experiment}

The crystals used are a cylindrical HEMEX-grade sapphire from Crystal Systems (USA), 5 cm diameter $\times$ 3 cm height, and containing 1-3 parts per million (ppm) of Fe$^{2+}$ and Fe$^{3+}$ impurity ions.\cite{McGuire2002, Benabid2000, Route2002} In the present work, we present results using two crystals, denoted C1 and C2, in which the Fe$^{3+}$ concentration is measured to be 10 and 100 part per billion (ppb), respectively.\cite{Creedon2009IEEE, Karim2010} The crystals are cut such that the $c$-axis of the sapphire lattice is parallel to the cylindrical $z$-axis. To attain the highest $Q$-factor at low temperature, the crystal is ordinarily mounted in the center of a cylindrical silver plated copper cavity.  Energy is coupled to the resonator through two straight antennas, placed 180$^{\circ}$ from each other, entering through the top of the cavity and oriented parallel to z-axis. The energy is stored in the crystal in high azimuthal order Whispering Gallery modes where the $\vec{H}$ field is predominantly oriented perpendicular to the $z$-axis. The crystal was cooled using a dilution refrigerator (DR), with a cooling power of 1 W at the 4 K stage, and $\sim$300 $\mu$W at 100 mK. A superconducting 5 Tesla magnet was mounted to the 4 K stage of the DR such that samples connected to the mixing chamber plate ($\sim$20 mK) could be placed in the bore of the magnet. A sapphire was mounted in the center of the magnet coil, with the walls of a radiation shield used to confine the energy of the WG modes rather than the usual cavity, which was too large to fit into the bore of the magnet. As a result, the $Q$-factors of the modes were strongly degraded and reached values around 10$^6$ instead of $10^8$ to $10^9$ which is typical with a high quality cavity. Measurements of the WG modes were performed using a network analyzer with an excitation power of \hbox{--40 dBm at the input of the probe antennae}. This power was low enough to allow measurements to be made without perturbing the temperature of the system. At higher power levels, the temperature of the sapphire rises quickly to $\sim$250 mK with an associated frequency shift. Using low levels of applied power also allowed the observation of absorption effects due to the presence of paramagnetic impurity ions, and allowed complex thermal bistability effects to be avoided, which have been observed in such a system in the past.\cite{Creedon2010temp} When the input power level is extremely low, all the energy inside the crystal is absorbed, making the observation of the ESR variations with the application of a magnetic field easy to do. For higher power measurements, the temperature was controlled at 400 mK.\\
As described earlier, the transitions of interest were $\left|\pm1/2\rangle\right. \rightarrow \left|\pm3/2\rangle\right.$. The choice of characterising these transitions only was due to several experimental limitations. The network analyzer used for the experiment was limited to measurements below 20 GHz, and in addition, the microwave transmission lines installed in the dilution fridge presented very high loss above 18 GHz due to their connector type. The cables were connected via feedthroughs between each temperature stage of the DR, with attenuators connected at each feedthrough to allow full thermalisation of the cables and reduce temperature gradients in the system. The crystals were characterized up to a magnetic field of 600 G, which corresponds to a frequency interval of 10--14 GHz for the ESR of Fe$^{3+}$. At higher magnetic field, more careful measurement must be made because of the complexity of the Zeeman spectra. The energy levels of different ions cross each other and it becomes difficult to identify them clearly without following the 19 GHz line upwards from 0 G. If we take in account ESR of other ions present in the sapphire such Cr$^{3+}$, Mn$^{4+}$ and V$^{+2}$,\cite{Geschwind1962, Brocherts1963} the system quickly becomes very complex to study, even if the concentration of these ions is very low. The magnetic field value was set manually through a magnet controller with a ramp time of about one minute between field values. After the set value was reached and the persistent current switch of the magnet closed, measurements were only performed after the temperature of the system had stabilized at the base temperature of $\sim$20 mK.

\begin{figure}[h!!!!!!]
\centering
\includegraphics[width=0.4\textwidth]{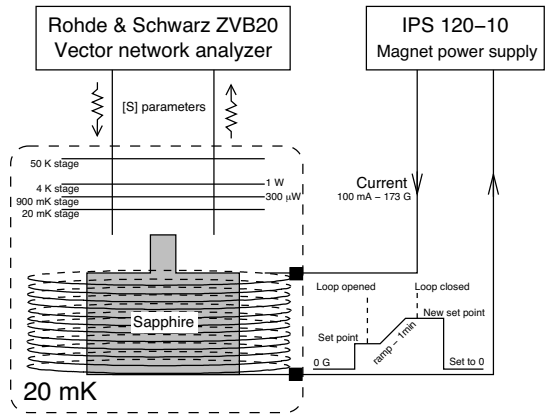}
\caption{\label{fig:setup2}Schematic of the experimental setup.}
\end{figure} 
\section{Measurement methodology}
The characterisation of the ions in our crystals is based on measurements of the AC susceptibility of the WG modes around the Fe$^{3+}$ ESR frequency when a DC magnetic field is applied parallel to the $c$-axis. The \hbox{S-parameters} of the WG modes between 10 and 14 GHz are measured, which act as filters sharply defined in frequency that allow the frequency of the ESR to be determined as the applied magnetic field is changed. When the frequency of the ESR is far from a WG mode, no change is observed. When the ESR frequency is coincident with a WG mode frequency, the S-parameters will decrease in amplitude, the frequency of the mode shifts, and the $Q$-factor of the mode is significantly degraded. This process is fully described in Section \ref{suscep} and related to the AC susceptibility of the ions.
\begin{figure}[h!!!!!!]
\centering
\includegraphics[width=8cm]{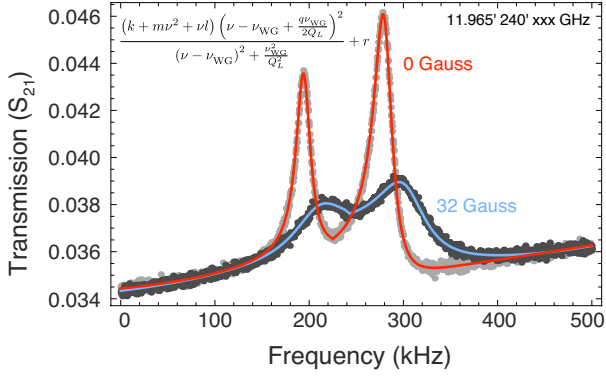}
\caption{\label{fig:example} Example traces of a WG mode showing the transmission S-parameter for zero applied DC magnetic field, and for a 32 Gauss applied field. The WG modes are degenerated and show generaly a double resonance due to cavity imperfections and alignements}
\end{figure} 
The lineshapes of the WG mode resonances are modelled with a Fano resonance fit. The equation of the fit is shown in Fig. \ref{fig:example}, and allows the mode frequency and $Q$ to be computed. From the results obtained, the AC susceptibility is calculated, and the physical parameters of the ions are determined.


\subsection{\label{suscep}AC magnetic susceptibility}

\begin{figure}[h!!!!!!]
\centering
\includegraphics[width=0.45\textwidth]{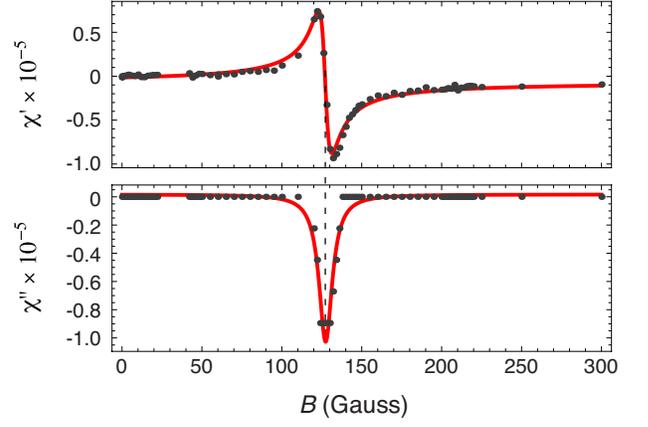}
\caption{\label{fig:chi}Real and imaginary parts of the magnetic susceptibility as a function of magnetic field strength for one of the WG mode doublet at 12.390 GHz.}
\end{figure}

The magnetic susceptibility of the ions in sapphire is described by the following general equation:
\begin{equation}
\chi(\nu)=\chi'(\nu)+i\chi''(\nu)
\end{equation}
We assume here the ESR is homogeneously broadened, and the resonance lineshapes are Lorentzian. As will be demonstrated in Section \ref{inhomog}, the ions are in fact inhomogeneously broadened at zero applied magnetic field. Here, we assume them to behave homogeneously when the Zeeman components of the ESR are split with the application of a magnetic field and are far away from each other in frequency. The analytical equation of the AC magnetic susceptibility is given by Kramers-Kr\"onig relationships:
\begin{equation}
\label{ac-suscep}
\begin{aligned}
\strut \chi'(\nu) &= \dfrac{(2\pi\tau_2)^2\nu_{ij}\chi_0(\nu-\nu_{ij}) }{1+(2\pi\tau_2)^2(\nu-\nu_{ij})^2+\dfrac{1}{4}\tau_1\tau_2(\gamma H)^2\sigma_{ij}^2} \\ 
\chi''(\nu) &=\dfrac{-2\pi\tau_2\nu_{ij}\chi_0}{1+(2\pi\tau_2)^2(\nu-\nu_{ij})^2+\dfrac{1}{4}\tau_1\tau_2(\gamma H)^2\sigma_{ij}^2}
\end{aligned}
\end{equation}
where, $\tau_1$ and $\tau_2$ are the spin-lattice and spin-spin relaxation times of the transition $ij$, respectively, 
$\sigma_{ij}^2=2$ for the transitions at 12 GHz, the term $\frac{1}{4}\tau_1\tau_2(\gamma H)^2\sigma_{ij}^2\ll1$, 
and the DC magnetic susceptibility is $\chi_0$. The real and imaginary parts of the magnetic susceptibility are related 
to the frequency shift and $Q$-factor of the modes as follows:
\begin{numcases}
\strut 
\dfrac{\nu_{WG}(B=0)-\nu_{WG}(B)}{\nu _B}=-\dfrac{1}{2}\eta \chi'\nonumber\\
\label{eqn:et}\\
\dfrac{1}{Q_{L}^{WG}(B)}-\dfrac{1}{Q_L^{WG}(B=0)}=\eta \chi''\nonumber
\end{numcases}
where, $\nu_{WG}$ and $Q_L^{WG}$ are the frequency and loaded $Q$-factor of the WG mode.
As an example of the fit (Fig.~\ref{fig:chi}) shows the real and imaginary parts 
of the magnetic susceptibility of one mode at $\nu_{WG}$=12.390 GHz.
At low magnetic fields, the Fe$^{3+}$ ESR is centred around a single frequency. When the magnetic field increases, the ESR splits into two Zeeman components with two resonant frequencies: $\nu_{low}$, which shifts to the lower frequencies, and $\nu_{high}$ to which shifts higher in frequency toward 12.390~GHz - the frequency of the WG mode considered here. The frequency of the mode, $\nu_{WG}$, will increase as $\nu_{high}$ approaches it, until it reaches a maximum and decreases to return to the initial value, at which point $\nu_{high}=\nu_{WG}$. When the magnetic field is further increased, $\nu_{high}$ continues to increase and the frequency of the WG mode decreases to a minimum before returning to its initial value. The $Q$-factor of the mode stays constant while $\nu_{high}\ll\nu_{WG}$, but as it becomes close, the losses due to the presence of the ions increases and the $Q$-factor of the modes decreases. When $\nu_{high}\gg\nu_{WG}$, the $Q$-factor then increases to reach it's initial value. The characterisation of all WG modes between 10 and 14 GHz allowed the measurement of the frequency dependence of each Zeeman component. Figure \ref{fig:sensitivityvsB} shows the evolution of the ESR frequency with magnetic field strength. The center frequency is defined as that where the real part of the susceptibility of the mode crosses zero, or when the imaginary part reaches it's minimum.


\subsection{ESR frequency and sensitivity to DC magnetic field}
\label{subsec:ESRvsB}

\begin{figure}[h!!!!!!]
\includegraphics[width=8.5cm]{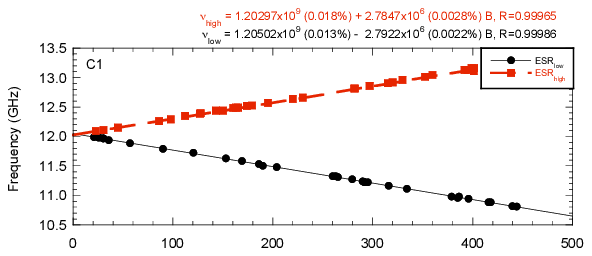}
\includegraphics[width=8.5cm]{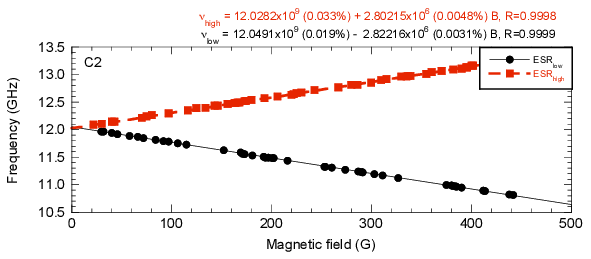}
\caption{\label{fig:sensitivityvsB}Frequency sensitivity to DC magnetic field for two sapphires at 20 mK. Only the points for a magnetic field higher than 20 Gauss are considered because of the complexity of the ions behavior (cf. section \ref{inhomog}).}
\end{figure}

As described earlier, the AC susceptibility allows the frequency of the Fe$^{3+}$ ESR to be determined when subjected to an applied DC magnetic field. By following the shift of the Zeeman components, it's possible to obtain their frequency sensitivity $S_{\text{high/low}}$ to the magnetic field, as shown by Fig. \ref{fig:sensitivityvsB}. The theoretical value is about 2.8 MHz/Gauss, assuming isotropic Zeeman splitting. However, the results of the fit show that this sensitivity is also dependent on the concentration of Fe$^{3+}$ ions in the sapphire. Indeed, for both crystals $S_{\text{high}}$ is found to be higher than $S_{\text{low}}$, and the sensitivity of the sapphire (C2) is higher than for the second crystal, C1. The theoretical prediction of 2.8 MHz/Gauss is obtained from the first order approximation of Eqns. \ref{eqn:energy} and \ref{eqn:fre} by using parameters shown in Table \ref{tab:Kornienko}. This fact highlights the novelty of this work because the characterisation of the paramagnetic ion in the past was performed with the traditional ESR method where a microwave signal is applied to the crystal in a microwave cavity at a fixed frequency near the ESR of the ion, and then a DC magnetic field is applied to the cold system. The measurement of the frequency shift allows the calculation of the real part of the ac susceptibility and then the estimation of the Hamiltonian parameters and sensitivities. In our case, we map out the path of the ESR frequency from zero field to the desired value of the magnetic field, giving a more accurate estimation of the Zeeman components frequency sensitivities. Our results reveal the anisotropic nature of these components, and we later give an estimation of the $g$-factor based on these results.

\begin{figure}[h!!!!!!]
\centering
\includegraphics[width=8.7cm]{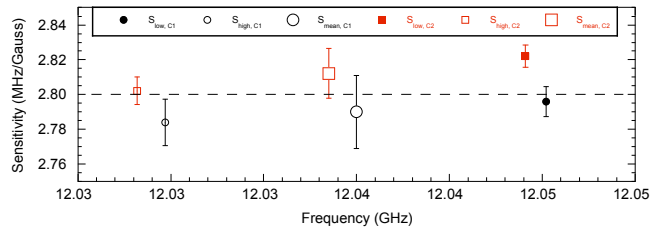}
\caption{\label{fig:sensitivit}ESR Zeeman component sensitivities as a function of frequency.}
\end{figure}

The frequencies as a function of DC magnetic field are given in Fig. \ref{fig:sensitivityvsB}. The fit shows that the zero-field ESR frequency is different for the upper and lower Zeeman components, which is consistent across both crystals. The crystal C2 contains has an ion concentration 10 times that of C1, and it has been shown in a previous publication how the greater concentration of ions revealed very complex behaviours. In the present case, the susceptibility obtained for low magnetic field ($B\le20$ Gauss) does not follow a Lorentzian shape and is not symmetric around the mode frequency, making the fit invalid and estimation of the correct resonance frequency difficult. As a first estimation, the central frequency is situated in the range 12.028-12.050 GHz. We expect that the two plus and minus spin packets do not perfectly overlap at zero magnetic field and form a large ESR.  The different sensitivities are show in Fig. \ref{fig:sensitivit} where the measured values are compared to the parameters in in Table \ref{tab:Kornienko}. The sensitivities are presented as a function of the ESR frequencies at zero field. It is clear that for both crystals the sensitivity of the ESR$_{\text{high}}$ is higher than the sensitivity of the ESR$_{\text{low}}$.  Also, the sensitivity of the crystal with a higher ion concentration is greater than for the lower concentration crystal.


\subsection{Determination of $g$-factor}

The measurement of $g$ is well known, and different techniques have been employed in the past for that purpose. A one-electron quantum cyclotron measurement is the most accurate, with a precision exceeding one part per trillion, measuring $g/2 = 1.001 159 652 180 85 (76)$.\cite{Aoyama2008, Odom2006} The determination of $g$ has been achieved in numerous materials and techniques, such GaAs, CdTe, InP, and graphite.\cite{RZhouGaMnAs, Singer1962, Stankowski2000, Pfeffer2012, Nibarger2003, Ermina1997} In Al$_2$O$_3$ crystals, electron spin resonance measurements have allowed the calculation of this factor for different ions in the microwave range. These measurements are summarised in Table \ref{tab:gfactor}. Other measurements of energy spectra in the optical and millimetre wave regimes have allowed the determination of this factor with great accuracy. For the Fe$^{3+}$ ion in sapphire, the $g$-factor determination is a poorly represented measurement in the literature.\cite{Luiten1996} 
\begin{figure}[h!!!!!!]
\centering
\includegraphics[width=8cm]{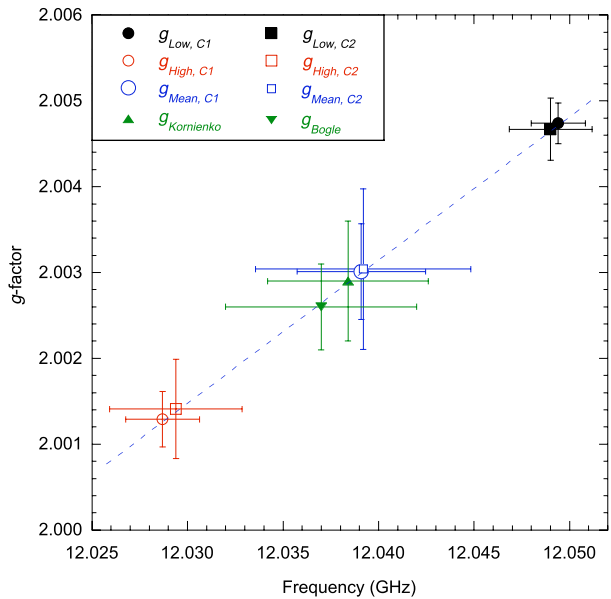}
\caption{\label{fig:landefit}Land\'e $g$-factor vs. frequency as measured in this work in two sapphires, C1 and C2, as well as the values measured by other authors.The $g-$factor is extracted from the fit of the WG modes frequency sensitivity to DC magnetic field as one parameter.}
\end{figure} 

\begin{table}[h!!!!!]
\caption{\label{tab:gfactor}Measured $g$-factor for a variety of impurity ions and host materials.\\}
\begin{tabular}{lcccccl}
\\
Material\hspace{20pt} 	&& Ion && 	$g$-factor															&& Reference\\ \hline\hline
Al$_2$O$_3$ 				&& Fe$^{3+}$ 			&&	2.0026$\pm$0.0005 											&& \cite{Symmons1962}\\
			  							&&			    				&&	2.0034$\pm$0.0003										&&\cite{Kornienko1961}\\ \hline
Al$_2$O$_3$ 				&& Mo$^{3+}$			&&	1.98$\pm0.01$ ($\perp$)								&&\cite{Sharoyan1974}\\
			  							&&				   				&&	1.968$\pm0.001$ ($\parallel$)						&&\\ \hline
Al$_2$O$_3$ 		 		&& Cr$^{3+}$			&&	1.9867$\pm0.0006$ ($\perp$)						&&\cite{DongPing2000}\\
			  							&& (ruby)					&&	1.9840$\pm0.0006$ ($\parallel$)					&&\\ \hline
TiO$_2$ 	  						&& Fe$^{3+}$ 			&&	2.000$\pm0.005$											&&\cite{David1960}\\ \hline
TiO$_2$ 	 			  			&& V$^{4+}$		    			&&	$g_x$=1.914														&&\cite{Gallay1986}\\ 
			 	 			  			&& 								&&	$g_y$=1.912														&&\\ 
			 	 			  			&& 				    			&&	$g_z$=1.956														&&\\ \hline
Emerald			 				&& Cr$^{3+}$			&&	1.97$\pm0.01$ ($\perp$)								&&\cite{Squire1966}\\
			  							&&				    			&&	1.973$\pm0.002$ ($\parallel$)						&&\\ \hline
\end{tabular}
\end{table}

Kornienko \& Prokhorov, and Symmons \& Bogle have estimated this parameter and claimed that $g$ is independent of temperature, while they found the other Hamiltonian parameters to be temperature dependent. This is in contradiction to other materials and ions where the $g$-factor is ordinarily temperature dependent,\cite{Ermina1997, Stankowski2000} thus we expect that the earlier measurements were not achieved with enough accuracy to reveal temperature induced variations. In addition, $g$ is expected to be isotropic for Fe$^{3+}$ in sapphire in the direction perpendicular to the crystal $c$-axis. In the present work, the $g$-factor is obtained from one parameter linear fit of the data points shown in Fig. \ref{fig:sensitivityvsB}. Figure \ref{fig:landefit} shows the $g$-factor as a function of the resonance frequency. For the two crystals, $g$ is smaller for the upper Zeeman component than for the lower component. The values are far from the those already estimated in the past. However the mean value of $g$ is close to that obtained by Bogle and Prokhorov.  In addition, it appears that the $g-$factor is dependent on the ion concentration. Where the concentration is higher, the $g-$factor is closer to 2.003 for both Zeeman components. The $g$-factor is also affected by an important parameter of the crystal, which is the alignment of the cylindrical $z$-axis with the crystal lattice $c$-axis. HEMEX-grade crystals show a tolerance of 1$^{\circ}$ that affects the $g$-factor by $\Delta g=0.00015$, which can be added to the fit tolerance quadratically.


\subsection{Magnetic losses due to ions}

\begin{figure}[h!!!!!!]
\centering
\includegraphics[width=8.5cm]{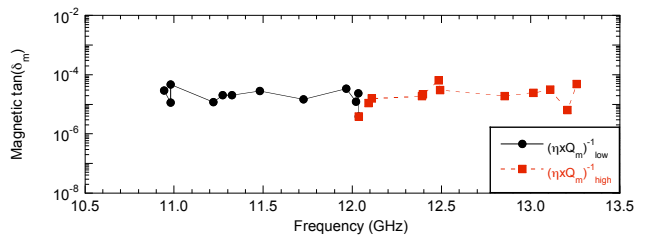}
\caption{\label{fig:maglosses}Magnetic losses due to Fe$^{3+}$ as a function of frequency. For clarity reasons, only the most relevant results of crystal C1 are presented. }
\end{figure}

The WG modes of the resonator are characterised by the $Q$-factor given by Eqn. (\ref{eqnQ}).\cite{Jerzy1999, Tobar2003, John2007} The loaded $Q$-factor of the mode ($Q_L^{\text{WG}}$) is related to the unloaded $Q$-factor ($Q_0$), which is a function of the external losses ($Q_e$ ) and the losses due to the presence of magnetic ions ($Q_{m}$).
\begin{equation}
\label{eqnQ}
\dfrac{1}{Q_L^{\text{WG}}} =\dfrac{1}{Q_0}-\dfrac{1}{Q_e}+\dfrac{1}{Q_{m}}
\end{equation}
The magnetic quality factor $Q_{m}$ is related to the magnetic losses in the resonator, and can be determined by the decreasing value of $Q_L^{\text{WG}}$ with the application of a magnetic field. When the frequency of the ESR does not match the mode frequency, then the magnetic losses are negligible ($Q_m^{-1}\rightarrow 0$). However, when the ESR frequency is tuned equal to the WG mode frequency, this magnetic loss mechanism becomes important and the $Q$-factor of the mode can be loaded significantly.
\begin{equation}
\dfrac{1}{Q_m} =\dfrac{1}{Q_L^{\text{WG}}(\nu_{_{\text{ESR}}}\neq \nu_{_{\text{WG}}})}-\dfrac{1}{Q_L^{\text{WG}}(\nu_{_{\text{ESR}}}= \nu_{_{\text{WG}}})}
\end{equation}
The magnetic $Q$ is related to the magnetic loss tangent $\tan(\delta_m)$ as follows
\begin{equation}
\dfrac{1}{Q_m}=\eta \tan(\delta_m)
\end{equation}
The $\tan(\delta_m)$, or magnetic loss tangent, is shown in Fig. \ref{fig:maglosses} as a function of frequency at a temperature of 20 mK. The losses due to the presence of the Fe$^{3+}$ ions are similar for both Zeeman components, and stay relatively constant around $2\times10^{-5}$.


\subsection{\label{inhomog}Inhomogeneous broadening effect}

At zero applied DC magnetic field, the lineshape broadening of the Fe$^{3+}$ ESR is not well known. Previous work has shown that the ions are inhomogeneously broadened, which has resulted in the observation of a number of nonlinear effects such as bimodal maser operation,\cite{Benmessai2008PRL} four-wave mixing,\cite{Creedon2012} the generation of combs and third harmonics,\cite{Creedon2012comb} and a gyrotropic effect.\cite{Benmessai2009} The interactions in the resonator occur between the ions and WG modes, and the spatial and frequency distribution of both are fundamental in the nonlinear processes. The ions behave as a number of spin packets with different ESR frequencies which overlap to sum to a single broad resonance. The greater the homogeneous broadening of the ions, the narrower the total ESR bandwidth. Thus, it is possible to find some spin packets in the crystal having the same configuration of frequency and linewidth, but located at different sites which make interaction possible with few of them. In the section \ref{subsec:ESRvsB} the frequency sensitivity of the WG modes to a DC magnetic field is developed. However, at small magnetic field no description is presented. The results of the fit showed in Fig. \ref{fig:sensitivityvsB} shows a crossing between the two lines different from zero Gauss, indicating a shift in the central frequency. For the two crystals the shift is 3.7 Gauss, equivalent to 10 MHz shift in the central frequency as active internal magnetic field is acting.


\section{Ion concentration and spin-spin relaxation time}
\begin{figure}[h!!!!!!]
\centering
\includegraphics[width=0.45\textwidth]{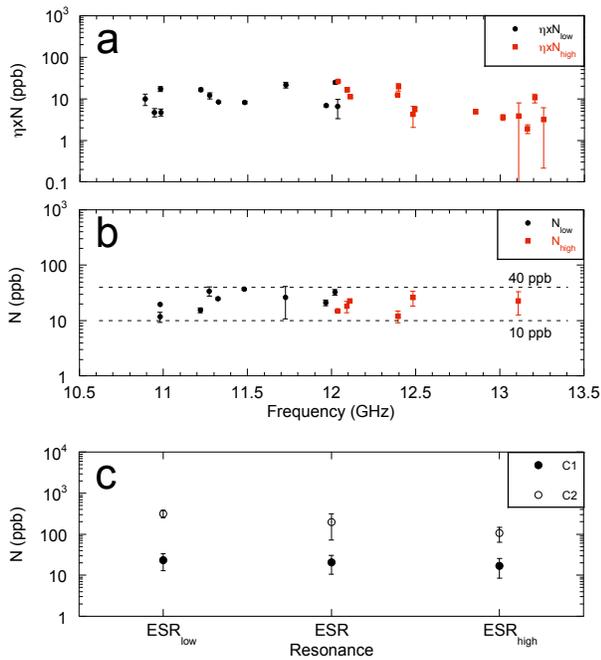}
\caption{\label{fig:ionconc} Ion concentration. (a) Shows $\eta\times N$ in units of ion/m$^3$, (b) shows the concentration in parts per billion, and (c) shows the total concentration calculated for all the modes. Only the results for the crystal C2 are shown for clarity.}
\end{figure}

The crystals characterized in the present work are HEMEX grade (grown using the heat exchange method) single-crystal $\alpha$-Al$_2$O$_3$. They contain extremely small trace concentrations of paramagnetic impurities due to the manufacturing process, such as Cr$^{3+}$, Mo$^{3+}$, Ti$^{3+}$ and Ti$^{4+}$, V$^{+}$, Mn$^{2+}$, Mn$^{3+}$ and Mn$^{4+}$, and Fe$^{+}$, Fe$^{2+}$, and Fe$^{3+}$. The concentration of each ion is approximately 1-2 ppm. It has been shown previously through measurements of the AC magnetic susceptibility via WG modes that the effective concentration of the ion of interest (Fe$^{3+}$) is about 10 parts per billion.  It has been shown that mass conversion of Fe$^{2+}$ ions into to Fe$^{3+}$ can be achieved by annealing the crystal in air, with the Fe$^{3+}$ concentration increasing by a factor of 10 in crystal C2 relative to C1. In parallel to these results it has been shown that the spin-spin relaxation time, related to interactions between nearby spins in the sapphire lattice, is about 2 ns - a value smaller than that found in the literature of $\sim$10 ns at 4.2 K. This relaxation time is related to the ESR linewidth, which is $\sim$27 MHz at 4.2 K. Shorter spin-spin relaxation times correspond with increased inhomogeneous broadening of the ESR linewidth. In this section, we characterise this parameter for our crystals as a function of the applied magnetic field. The ion concentration and spin-spin relaxation time can be estimated for each crystal from the AC magnetic susceptibility defined by Eq. \ref{ac-suscep}. The concentration is extracted from the DC part of the magnetic susceptibility as follows:
\begin{equation}
\chi_0=\dfrac{(g\mu_B)^2\mu_0}{2h\nu_{ij}}\Delta N_{ij}\sigma_{ij}^2
\end{equation}
where $\Delta N_{ij}$ is the population difference between levels $i$ and $j$, which is determined by a Boltzmann distribution. By considering $N_{ij}$, the total population of the transition between $i$ and $j$, we find
\begin{equation}
\Delta N_{ij}=\tanh \left(\dfrac{g\mu_B B_{ij}}{k_B \textrm{T}}\right) N_{ij}
\end{equation}
It is possible to link the frequency $\nu$ to the corresponding magnetic field by a first order approximation of Eq. \ref{eqn:fre} in the limit $B<3000$ G:
\begin{numcases}
\strut \nu \equiv \nu_{B=0} \pm S\times B\nonumber\\
\\
\nu_{ij} \equiv \nu_{B=0} \pm S\times B_{ij}\nonumber
\end{numcases}
Substituting $\Delta N_{ij}$ into the Kramers-Kr\"onig relationships gives
\begin{numcases}
\strut \chi'(B)= \dfrac{a ^2\times b\times S (B-B_{ij})}{1+(a \times S)^2(B-B_{ij})^2}\nonumber\\
\\
\chi''(B)=-\dfrac{a\times b}{1+(a \times S)^2(B-B_{ij})^2} \nonumber
\end{numcases}
where
\begin{align}
\strut a &=2 \pi \tau_2\text{,\hspace{1pc}and}\\
b &=\dfrac{(g\mu_B)^2}{2h} \mu_0\sigma_{ij}^2N_{ij} \tanh \left[\dfrac{h}{2 k_B\textrm{T}} (\nu_{B=0}-S B_{ij})\right]
\end{align}
Note that the parameter $a$ defines the spin-spin relaxation time. From equation \ref{eqn:et} it is essential to determine the correct filling factor for each WG mode to extract the concentration. Due to the number of modes used in the fit it is necessary to determine the filling factors using a rigorous electomagnetic simulation of the Whispering Gallery modes.\cite{Lefloch2008APL} Because the density of modes is high and the simulation process lengthy, the determination of the correct modes is often difficult. As a result, we extract the product $\eta N$ from the real part of the AC susceptibility, then normalise the product to the highest value obtained from the fit. The result of the fit which gives $\eta N$ is shown in Fig. \ref{fig:ionconc}.

The error bars in the figure are statistical errors from the fit results. They are due to the number of points around the mode frequency, and to the method of interaction with the mode. The spatial distribution of some WG modes does not encompass a large enough quantity of ions which makes the detection of the ESR frequency difficult. Nonetheless, the results give a good approximation to the value of $\eta$ with the assumption that a similar number of ions interact with all WG modes. The filling factors extracted are then used in the imaginary part of the AC susceptibility $\chi''$ to compute the corrected $N$ and $\tau_2$. The WG modes whose $Q$-factor degraded by 10\% or greater when the ESR frequency was coincident with the WG mode frequency were selected in the fit (Fig. \ref{fig:ionconc}). For the crystal C1, the concentration is in the range of 10 to 40 ppb, and for C2 it is in the range 100 to 300 ppb. The total concentration should be the same for both Zeeman components, as the only difference is the number of ions seen by each mode. Therefore the number of ions is fixed by the WG mode volumes. For example, the mode at 12.04 GHz encloses a volume of about $11.7\times10^{-6}$ m$^3$, which contains a population of of $5.6\times 10^{15}$ ions for the crystal C1, and $5.4\times 10^{16}$ ions for the crystal C2. Figure \ref{fig:ionconc} shows the estimated concentrations for the two crystals.

The spin-spin relaxation time is deduced from the fit in the same way, and the results are summarised in Table \ref{tab:tau2vsF}. A comparison between the real and imaginary parts of the AC magnetic susceptibility is made. The results are similar to each other and are in close agreement with a values found in the literature.

\begin{figure}[h!!!!!!]
\includegraphics[width=0.45\textwidth]{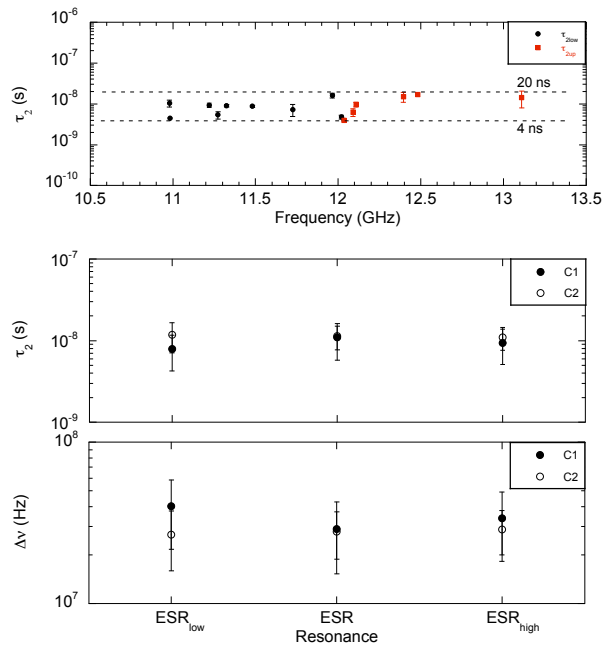}
\caption{\label{fig:tauvsF2}Spin-spin relaxation time $\tau_2$ as function of frequency. The corresponding ESR bandwidth is also shown.}
\end{figure}

\begin{table}[h!]
\caption{\label{tab:tau2vsF}Evaluation of the spin-spin relaxation time $\tau_2$. The evaluations for $\chi''$ were made taking in account the filling factor $\eta$. For comparison, the values given by Bogle \& Symmons\cite{Symmons1962} and Kornienko \& Prokhorov\cite{Kornienko1961} are given.}
\begin{tabular}{lcccccr}
\multirow{2}{*}{} & & \multicolumn{2}{c}{C1}& & \multicolumn{2}{c}{C2} \\ \cline{3-4} \cline{6-7}
& & $\tau_2$ (ns) & $\Delta \tau_2$ (ns) & & $\tau_2$ (ns) & $\Delta \tau_2$ (ns) \\ \hline \hline
\multirow{3}{*}{$\chi'$ fit} &ESR$_{\text{low}}$ &9.6 & 3.6 & &12.8 & 2.1\\
&ESR$_{\text{high}}$ &4.1 & 2.7 & &7.5 & $-$ \\
&ESR &6.9 & 4.2 & &11.0 & 3.2\\ \hline
\multirow{3}{*}{$\chi''$ fit} &ESR$_{\text{low}}$ & 8.1 & 3.5 & &11.9 & 4.8\\
&ESR$_{\text{high}}$ & 14.8 & 9.5 & &11.0 & 3.4\\ 
&ESR & 10.6 & 7.0 & &11.4 & 3.7\\ \hline\hline
& & $\tau_2$ (ns) & $\Delta \tau_2$ (ns) & & & \\ \hline \hline
\multicolumn{2}{l}{Bogle \& Symmons}& 11.8 & 2.2 \\
\multicolumn{2}{l}{Kornienko \& Prokhorov} & 11.8 & 2.4\\ \hline
\end{tabular}
\end{table}
Another way of determining $\tau_2$ is the width of the resonance defined by Eq. \ref{eqmn:deltanu}:
\begin{equation}
\label{eqmn:deltanu}
\Delta \nu=\dfrac{1}{\pi \tau_2}
\end{equation}
The bandwidth of each Zeeman component, and the spin-spin relaxation time of the ions are shown in Fig. \ref{fig:tauvsF2}. It is clear that the value of 30 MHz is very close to the one published by Bogle \& Symmons, and Kornienko \& Prokhorov (see Table \ref{tab:tau2vsF}). However, some work has shown that the ESR bandwidth is concentration dependent and cannot be accurately measured when highly doped ($N\gg$ 1 ppm).\cite{Buzare2002}

\section{Conclusion}
In conclusion, the physical parameters of Fe$^{3+}$ ions in high-purity sapphire have been determined at 20 mK by measuring the AC susceptibility of the ions using Whispering Gallery modes. A magnetic field is applied to split the ESR of the ion under study into two Zeeman components. The characterisation of the WG modes allowed the frequency evolution of each Zeeman component with magnetic field to accurately be determined. We measure the frequency sensitivities as $(2.81\pm0.022)$~MHz/Gauss for the crystal with high Fe$^{3+}$ concentration, and $(2.79\pm0.014)$~MHz/Gauss for the crystal with lower concentration. The zero field splitting has been measured as 12.039 GHz for both crystals. The difference in terms of frequency sensitivity can be explained by an anisotropic behaviour of the crystal depending on the concentration of ions involved in the process. However, calculations of the $g$-factor gives results similar to those found in the literature. It is expected that the Hamiltonian parameters do not change significantly between 4.2 K and 20 mK. Indeed, the behaviour of the ESR remains constant for temperatures below 4.2 K. To estimate these parameters for our crystals it is necessary to know the behaviour with magnetic field of the transitions at 19.3 and 31.3 GHz. Finally, the concentrations estimated are similar to the ones calculated in preceding publication using different methods, such the AC susceptibility measured by saturating the ions at 12.04 GHz, or calculated from a Whispering Gallery mode maser oscillator. For the crystal C2, a concentration of $\sim$200 ppb of Fe$^{3+}$ was measured, and for C1 a concentration of 20 ppb. In previous works, the ESR bandwidth of Fe$^{3+}$ in sapphire has been assumed to be around 27 MHz at 12.04 GHz. Here, we measure a value of $34 \pm 5.5$ MHz for  crystal C1 at $12.040\pm0.0063$ GHz, and a value of $29 \pm 9$ MHz for crystal C2 at $12.039\pm0.0038$ GHz. Within the bounds of experimental error, the bandwidth appears to be independent of the ion concentration.

\begin{acknowledgments}
This work was funded by Australian Research Council grant numbers DP1092690, DP0986932, CE110001013, and FL0992016, and Research Discovery Award grant number 12104411.
\end{acknowledgments}

\providecommand{\noopsort}[1]{}\providecommand{\singleletter}[1]{#1}%

\end{document}